\documentclass[12pt]{iopart}
\usepackage{graphicx}
\begin{document}

\title[Collective excitations of BEC under anharmonic trap position
jittering] {Collective excitations of BEC under anharmonic trap
position jittering}
\author{F Kh Abdullaev, R M Galimzyanov
\footnote [7]{To whom correspondence should be addressed
(ravil@uzsci.net)} , Kh N Ismatullaev}
\address{Physical-Technical Institute of the Academy of
Sciences, 700084, Tashkent-84, G.Mavlyanov str.,2-b, Uzbekistan}

\begin{abstract}
Collective excitations of a Bose-Einstein condensate under periodic
oscillations of a quadratic plus quartic trap position has been
studied. A coupled set of variational equations is derived for the
width  and the condensate wave function center. Analytical expressions
for the growth of oscillation amplitudes in the resonance case are
derived. It is shown that jittering of an anharmonic trap position can
cause double resonance of the BEC width and the center of mass
oscillation in the wide range of the BEC parameters values. The
predictions of variational approach are confirmed by full numerical
simulations of the 1D GP equation.
\end{abstract}

\pacs{03.75.Kk;67.40.Db;03.65.Ge}

\maketitle


\section{Introduction}
The investigation of low energy collective excitations is
important for understanding of dynamics of the atomic quantum fluids
(see the review \cite{Stringari}). Most of such theoretical and
experimental studies has been performed for the condensate trapped in a
harmonic (parabolic) trap. The description of the wavefunction dynamics
in such a trap has many simplifying properties both for repulsive and
for attractive interactions between atoms. The theory is based on the
Gross-Pitaevskii equation, which is the nonlinear Schr\"odinger
equation with the linear oscillator potential. The analysis for the
repulsive condensate shows that in such potential the motion of the
center of mass of condensate is decoupled from the oscillations of the
condensate width. This observation is also valid for the case of
attractive  BEC, where in a quasi 1D geometry the matter wave soliton
can exist. For the solitonic  wave in a parabolic potential, the
center-of-mass motion is well known to be completely decoupled from the
internal excitations and it represents the analog of the Kohn theorem
for solitonic wave packet \cite{Kohn}. It can be shown both on
the level of the symmetries of 1D GP equation and using the moments
method \cite{AbdGal}. The separate resonances in the soliton width and
in the position has been investigated in \cite{Baiz}. Resonances by the
periodic variation of the scattering length has been considered for 1D
Bose gas in \cite{AbdGar}.

For the case of an anharmonic trap potential the evolution of
translational mode (motion of the center of mass) and the internal mode
(oscillation of the width) becomes coupled. It gives rise to the
possibility to control internal modes by manipulating the position of
the trap. Such possibility may be useful also in creation of new
technological devices, including quantum computers \cite{Rolston} and
ultra sensitive interferometers \cite{Kasevich}. Frequencies for
low-energy excitation modes of a one-dimensional Bose-Einstein
condensate with repulsive interaction between atoms in a quadratic plus
quartic trap have been calculated in \cite{Li}.
An approximate solution to describe the dynamics of Bose- Einstein
condensates in anharmonic trapping potentials based on scaling
solutions for the Thomas- Fermi radii has been presented in \cite{Ott}.

In this work we study effect of periodic oscillations of the anharmonic
elongated trap potential position on the dynamics of BEC confined in
this trap. We will consider the case when the anharmonicity and the
oscillations amplitude are small.

\section{The model}
The dynamics of a trapped quasi-one dimensional Bose gases in
elongated in the longitudinal direction anharmonic trap can be
described in the framework of the 1D Gross-Pitaevskii equation
\begin{equation}\label{gpe}
i\hbar\phi_{t} = -\frac{\hbar^2}{2m}\phi_{xx}
+V(x,t)\phi + g_{1D}|\phi|^{2}\phi,
\end{equation}
with the total number of atoms $N = \int |\phi|^2 dx$. This equation is
obtained in the case of a highly anisotropic external potential under
the assumption that the transversal trapping potential is harmonic:
$V(y,z) = m\omega_{\perp}^{2}(y^2 + z^2)/2 $ and $\omega_{\perp} \gg
\omega_{x}.$  Under such conditions we can seek the solution of a 3D
equation in the form $U(x,y,z;t) = R(y,z)\phi(x,t)$ where $R_{0}^2 =
m\omega_{\perp}\exp(-m\omega_{\perp} \rho^2/\hbar)/(\pi\hbar)$.
Averaging in the radial direction (i.e. integrating over the
transversal variables) we have equation (\ref{gpe}) for the dynamics of
the gas in longitudinal direction. The effective one dimensional mean
field nonlinearity coefficient $g_{1D} = 2\hbar a_{s}\omega_{\perp}$,
with $a_{s}$ is the atomic scattering length. $a_{s} > 0$ corresponds
to the Bose gas with a repulsive interaction between atoms and $a_{s} <
0$ to attractive interaction.

The dimensionless form of equation (\ref{gpe})
\begin{equation}\label{gpe1}
i\psi_{t} + \frac{1}{2}\psi_{xx} - V(x,t)\psi - g|\psi|^{2}\psi = 0,
\end{equation}
can be obtained by setting:
$$t = \omega_{x}t,\ l = \sqrt{\frac{\hbar}{m\omega_{x}}},\ x= \frac{x}{l},
\ \psi = \sqrt{2|a_{s}|\omega_{\perp}/\omega_{x}}\phi,$$ with $g = \pm
1$ for the repulsive and attractive two-body interactions respectively.

\section{Variational analysis}
To describe the collective oscillations  of Bose gas in an anharmonic
trap we employ the variational approach. For this purpose, we use the
gaussian trial function for the wavefunction $\psi(x,t)$

\begin{eqnarray}\label{3}
\psi(x,t)=A(t)exp(-\frac{(x-x_0(t))^2}{2\eta^2(t)}+k(t)(x-x_0(t))+\nonumber\\
+\frac{ib(t)(x-x_0(t))^2}{2}+i\varphi(t)),
\end{eqnarray}
where A, $\eta$, b, $x_0$ and $\varphi$ are the amplitude, width,
chirp, center of mass and linear phase, respectively. The trap
potential is chosen of the form $V(x) = V_2 (x-c(t))^2 + V_4
(x-c(t))^4$, where c(t) is an external parameter describing forced
motion of the center of the trap.

Using this ansatz in obtaining the Euler-Lagrange equations we come to
the following system of equations for the width and the center of mass
of the wave packet
\begin{eqnarray}\label{sys1}
\eta_{tt} &=& \frac{1}{\eta^3} -2 \eta V_2 - 6 V_{4} \eta^{3} - 12
V_{4} \eta (x_0-c)^2 + \frac{g N }{\sqrt{2 \pi}\eta^{2}},
\end{eqnarray}
\begin{eqnarray}\label{sys2}
 x_{0tt} &=& -
2 V_2 (x_0-c) - 6 V_{4} \eta^{2} (x_{0} -c) - 4 V_{4}( x_{0}-c)^{3}.
\end{eqnarray}

Linearizing (\ref{sys1}) and (\ref{sys2}) around the equilibrium points
($\eta_{tt}=0, \ x_{0tt}=0$) we get the following set of equations
\begin{eqnarray}\label{linsys1}
\delta_{tt}=-w_\eta^2 \delta-12 V_4 \eta_s (x_0-c)^2 ,
\end{eqnarray}
\begin{eqnarray}\label{linsys2}
x_{0tt}=-w_x^2 (x_0-c),
\end{eqnarray}
where $\eta_s$ is the equilibrium point of the width,
$\delta=\eta-\eta_s$ is the deviation from the equilibrium point,
$w_{\eta}$ and $w_{x}$ are determined by expressions
\begin{eqnarray}\label{eigenfreqcoeff}
w_\eta^2 =  2V_{2} + 18V_{4}\eta_s^2 + \frac{3}{\eta_s^4} +\frac{\sqrt{2} g N}{\sqrt{\pi}\eta_s^3}
+ 12 V_4 x_s^2 , \nonumber\\
w_x^2 = 2V_{2}+6V_{4}\eta_s^2 + 12 V_4 x_s^2 .
\end{eqnarray}

For the excitation frequencies we have
\begin{eqnarray}\label{eigenfrequency}
w_{1,2} &=& \left(\frac{w_\eta^2+w_x^2+-\sqrt{(w_\eta^2-w_x^2)^2+4k_1
k_2}}{2}\right)^\frac{1}{2},
\end{eqnarray}
where
\begin{eqnarray}\label{k1k2}
 k_{1} = - 24 V_{4} \eta_s x_{s},\nonumber\\
 k_{2} = - 12 V_{4} \eta_s x_s.
\end{eqnarray}

Taking into account that in the equilibrium point $x_s=0$ and $w_\eta
>w_x$ we get

\begin{eqnarray}\label{eigenfrequency1}
w_1 = w_\eta = \sqrt{\frac{\sqrt{2} g N}{\eta_s^3} + 2V_{2} +
18V_{4}\eta_s^2 +
\frac{3}{\eta_s^4}}, \nonumber\\
w_2 = w_x = \sqrt{2V_{2}+6V_{4}\eta_s^2}.
\end{eqnarray}

\section{Resonance}

Let us suppose oscillation of the trap position to be periodical, viz
$c(t) = h sin(wt)$, where $w$ is the oscillation frequency. As easily
seen from the linearized equations, the center of mass and width
oscillations behave like periodically driven oscillator with the
"external forces" $c^2(t)$ in equation (\ref{linsys1}) and $c(t)$ in
equation (\ref{linsys2}). This means that the frequency of the
"external force" is equal to $2w$ in the first equation and to $w$ in
the second. Then a double resonance in oscillations of the center of
mass and the width is possible when $w=w_x=w_\eta/2$.

To describe the resonance growing of the width and center of mass
oscillations in equations (\ref{linsys1}) and (\ref{linsys2}) we seek
$\delta$ and $x_0$ as $\delta = A(t) sin(2wt+\phi_1)$ and $x_0=B(t)
sin(wt+\phi_2)$. Supposing that $A(t)$ and $B(t)$ weakly depends on
time and substituting these expressions into (\ref{linsys1}) and
(\ref{linsys2}) and assembling coefficients of $sin(2wt+\phi1)$,
$cos(2wt+\phi1)$, $sin(wt+\phi2)$ and $cos(wt+\phi2)$ we come to the
following differential equations for $A(t)$ and $B(t)$
\begin{eqnarray}\label{difeqForAandB}
A_t = \frac{3V_4\eta_s}{w} B^2 , \nonumber\\
B_t = \frac{hw}{4} + \frac{12 V_4 \eta_s}{w} A B.
\end{eqnarray}
For small amplitudes $|A|<<1$ and $|B|<<1$ we have the expressions
\begin{eqnarray}\label{eqForAandB}
A(t) = \frac{1}{16}V_4 \eta_s h^2 w t^3 , \nonumber\\
B(t) = \frac{hw}{4}t ,
\end{eqnarray}
which describe the growth of oscillation amplitudes.

\section{Numerical simulations}
We have carried out a series of time dependent simulations of the
system evolution based on the variational approach using
equations~(\ref{sys1}) and (\ref{sys2}) as well as exact numerical
computations of the full Gross-Pitaevsky (GP) equation (\ref{gpe1}). In
our numerical simulations of the GP equation (\ref{gpe1}) we discretize
the problem in a standard way, with the time step $dt$, and spatial
step $dx$, so $\psi^k_j$ approximates $\psi(jdx, kdt)$. More
specifically we approximate the governing equation (\ref{gpe1}) with
the semi-implicit Crank-Nickolson scheme using split-step method
\cite{Adh}. The results of numerical simulations of both PDE and ODE
models are presented below. In all ODE and PDE simulations the norm of
the BEC wave packet is taken to be $N = 1$.

\begin{figure}[h]
\centerline{\includegraphics[width=8cm,angle=-90]{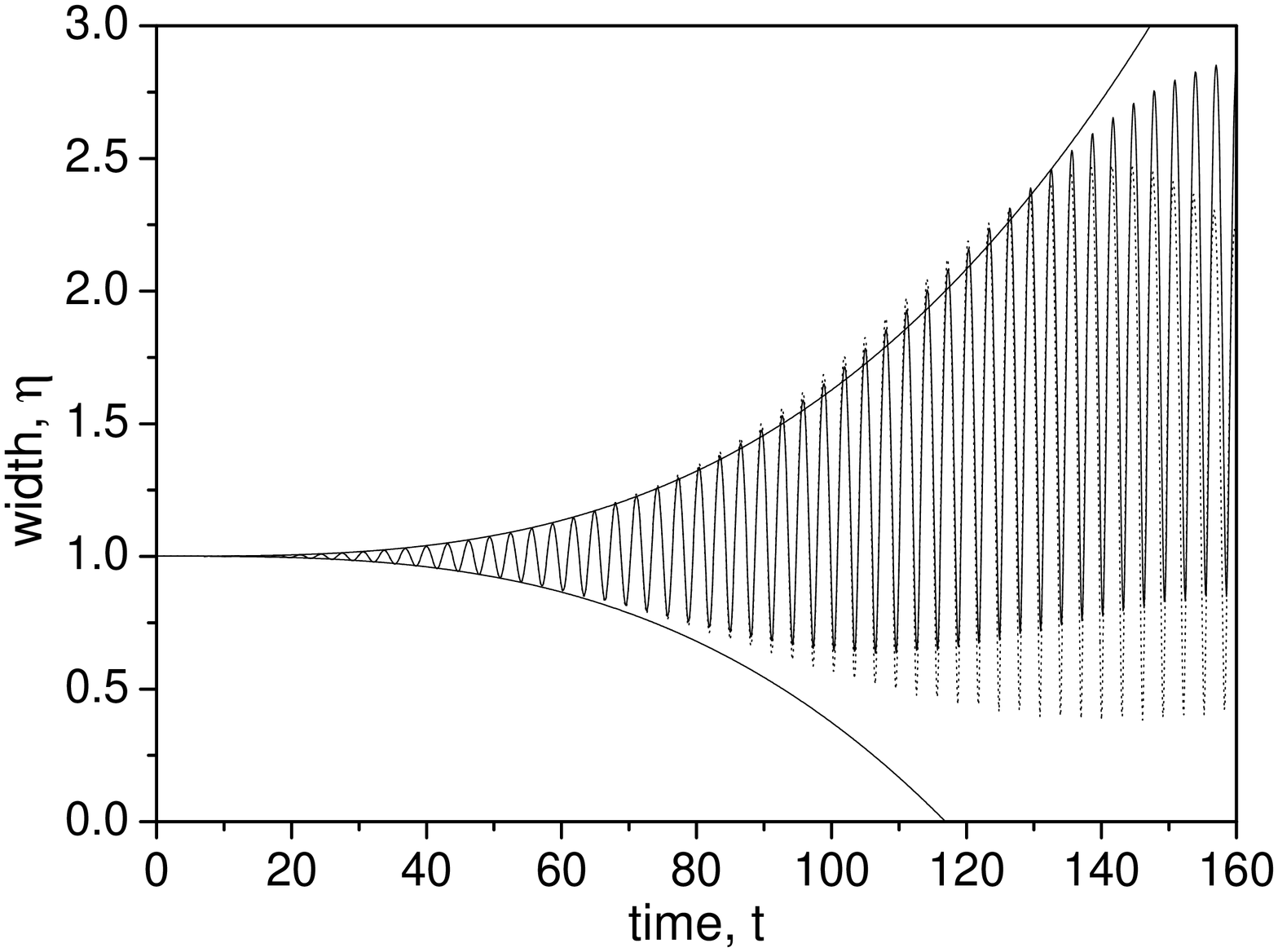}}
\centerline{\includegraphics[width=8cm,angle=-90]{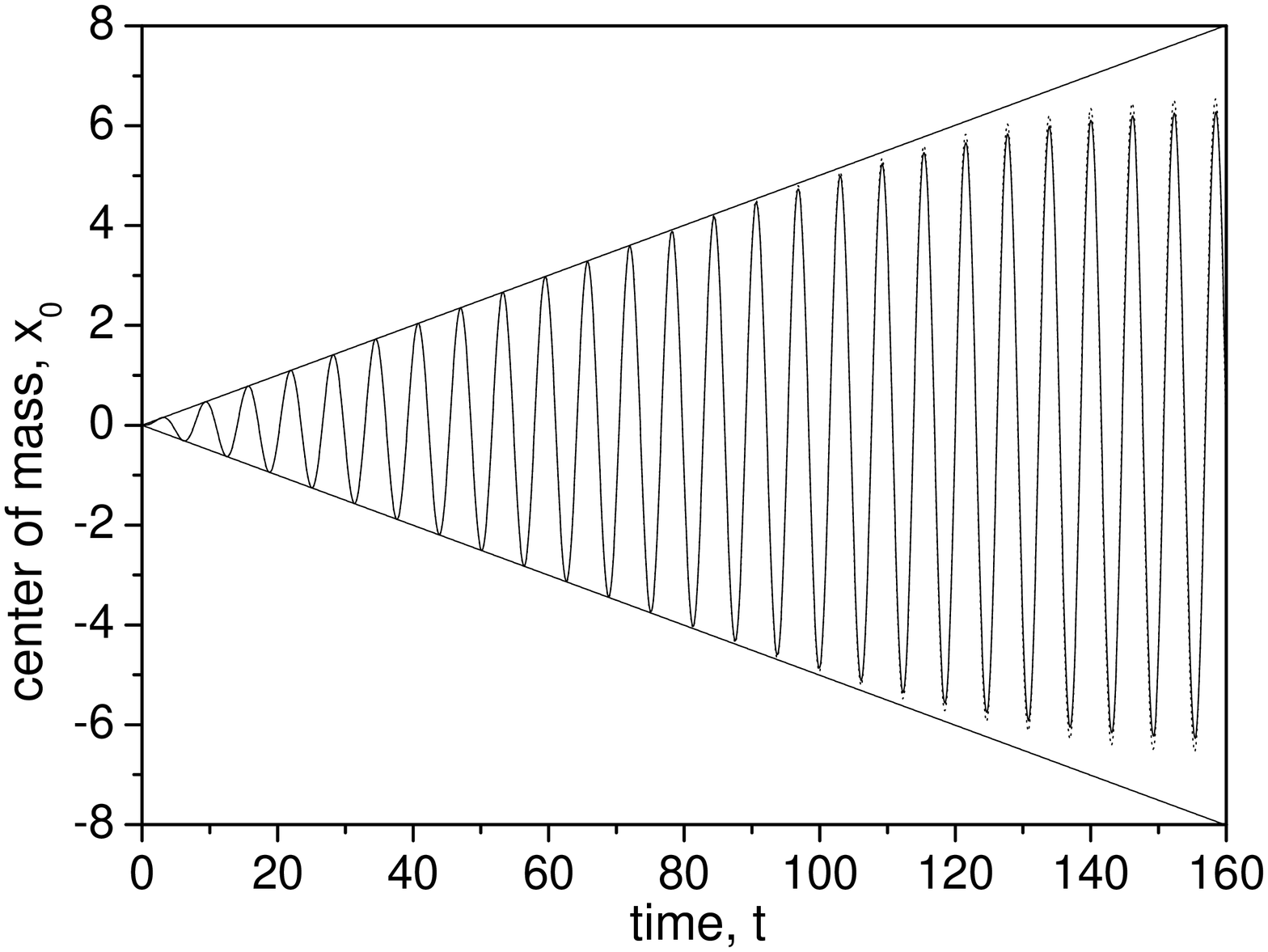}}
\caption{Resonance in oscillations of the width and the mass center of
the repulsive BEC in an anharmonic trap with the forced periodical
oscillation of the trap center by law $c(t) = h sin(w t)$.  The
parameters are $V_2 = 0.5$,\ $V_4 = 0.0005,\ h = 0.1$, $w_{\eta}/w_x =
2$ and $w = w_x$. The skirting lines present  theoretical prediction of
amplitude oscillations growth in the resonance. Solid and dotted lines
stand for PDE and ODE simulations respectively.} \label{figcenter}
\end{figure}

Figure \ref{figcenter} depicts double resonance in   oscillations of
the width and center of mass of the repulsive condensate. In PDE
simulations the initial
wave packet is taken in the ground state. The trap center position
oscillates in time periodically as $c(t) = h sin(w t)$ with the
amplitude $h = 0.1$. The parameters of the trap potential are $V_2 =
0.5$ and $V_4 = 0.0005$. The value of the forced oscillation frequency
$w = w_x$ at the condition $w_x=w_\eta/2$. Necessary value of the
nonlinearity coefficient $g$ providing this condition is obtained by
solving a set of equations
\begin{eqnarray}\label{resfreq}
w_x=w_\eta/2 , \nonumber\\
\frac{1}{\eta_s^3} -2 \eta_s V_2 - 6 V_{4} \eta_s^{3} + \frac{g N
}{\sqrt{2 \pi}\eta_s^{2}} &=& 0.
\end{eqnarray}
Here the second equation determines the equilibrium point of equation
(\ref{sys1}).

As seen, unlike the harmonic case, in an anharmonic trap potential the
forced oscillations of the trap center position induce oscillations not
only in the condensate center of mass but also in the condensate width.
In the figure for comparison full GPE and ODE simulations of the width
and center of mass oscillations are shown. Theoretical prediction is
shown by skirting lines described by equation (\ref{eqForAandB}).

Double resonance presented in figure \ref{figcenter} occurs under the
condition $w_x=w_\eta/2$. It corresponds to the particular value of the
nonlinearity coefficient $g = 0.015$ (repulsive BEC). However here we
meet with a very remarkable fact that in a wide range of the
nonlinearity values the ratio $w_\eta/w_x$ is close to 2.

\begin{figure}[h]
\centerline{\includegraphics[width=8cm,angle=-90]{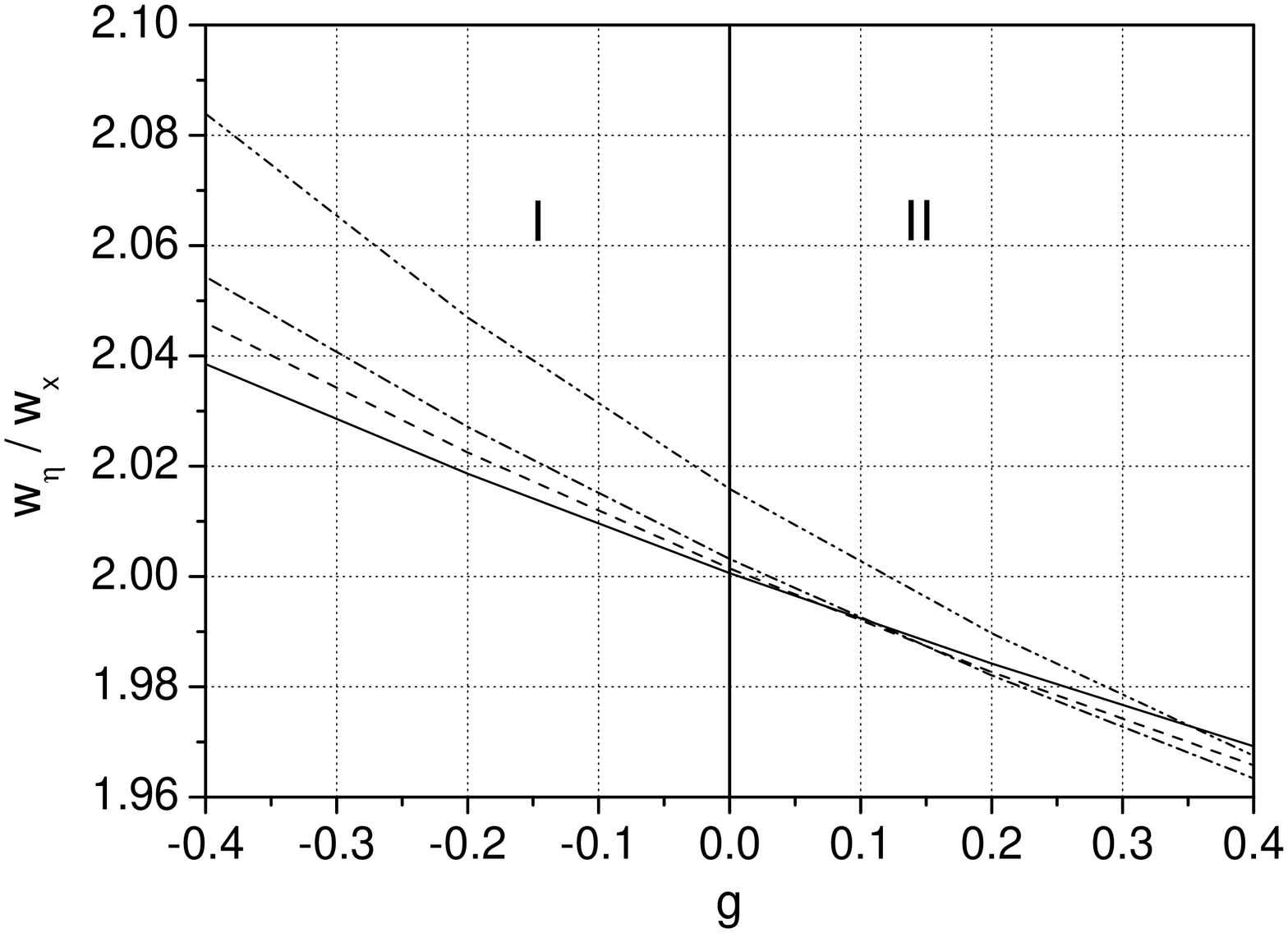}}
\caption{Ratio of eigenfrequencies of the BEC width and center of mass
oscillations  versus the nonlinearity $g$ for different values of
$V_2$. Solid, dashed, dot-dashed and dot-dot-dashed lines stand for the
cases $ V_2 = 0.9,\ 0.5,\ 0.3,\ 0.1$ respectively. Field $I$
corresponds to attractive  BEC and field $II$ to repulsive one}
\label{wawx}
\end{figure}

For confirmation it, in figure \ref{wawx} the ratio of eigenfrequencies
values of the width and mass center oscillations versus the
nonlinearity coefficient, $g$ is shown for several values of the
quadratic part of potential $V_2$. When $g$ ranges from -0.4
(attractive BEC) to 0.4 (repulsive BEC) the ratio changes from 2.084 to
1.964. Closeness of the ratio value
$w_{\eta}/w_x$ to 2 makes possible the existence of double resonance in
a very wide range of the Bose-Einstein condensate parameters under
oscillations of the anharmonic trap potential position at the frequency
$w = w_x$.

\begin{figure}[h]
\centerline{\includegraphics[width=8cm,angle=-90]{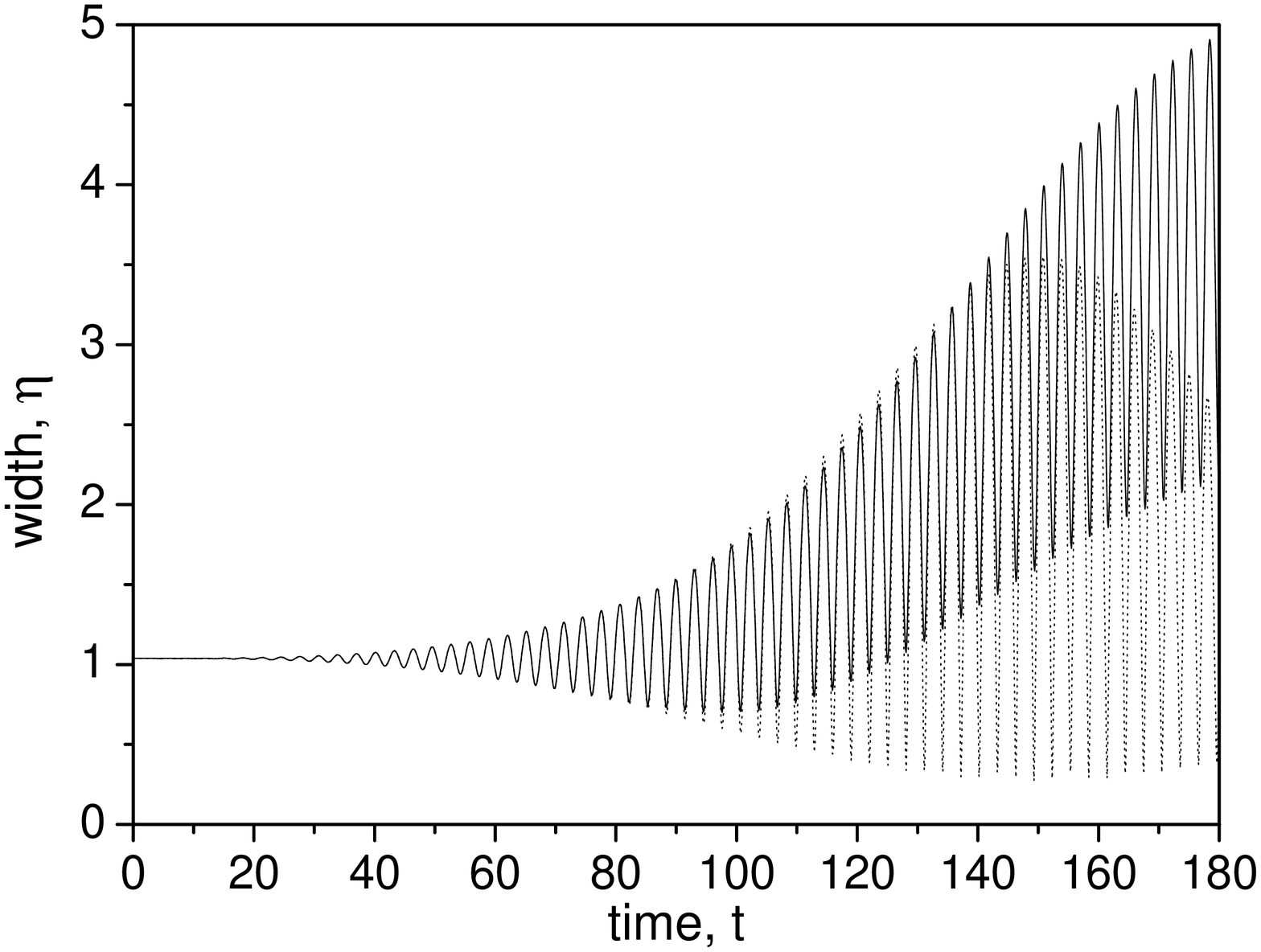}}
\centerline{\includegraphics[width=8cm,angle=-90]{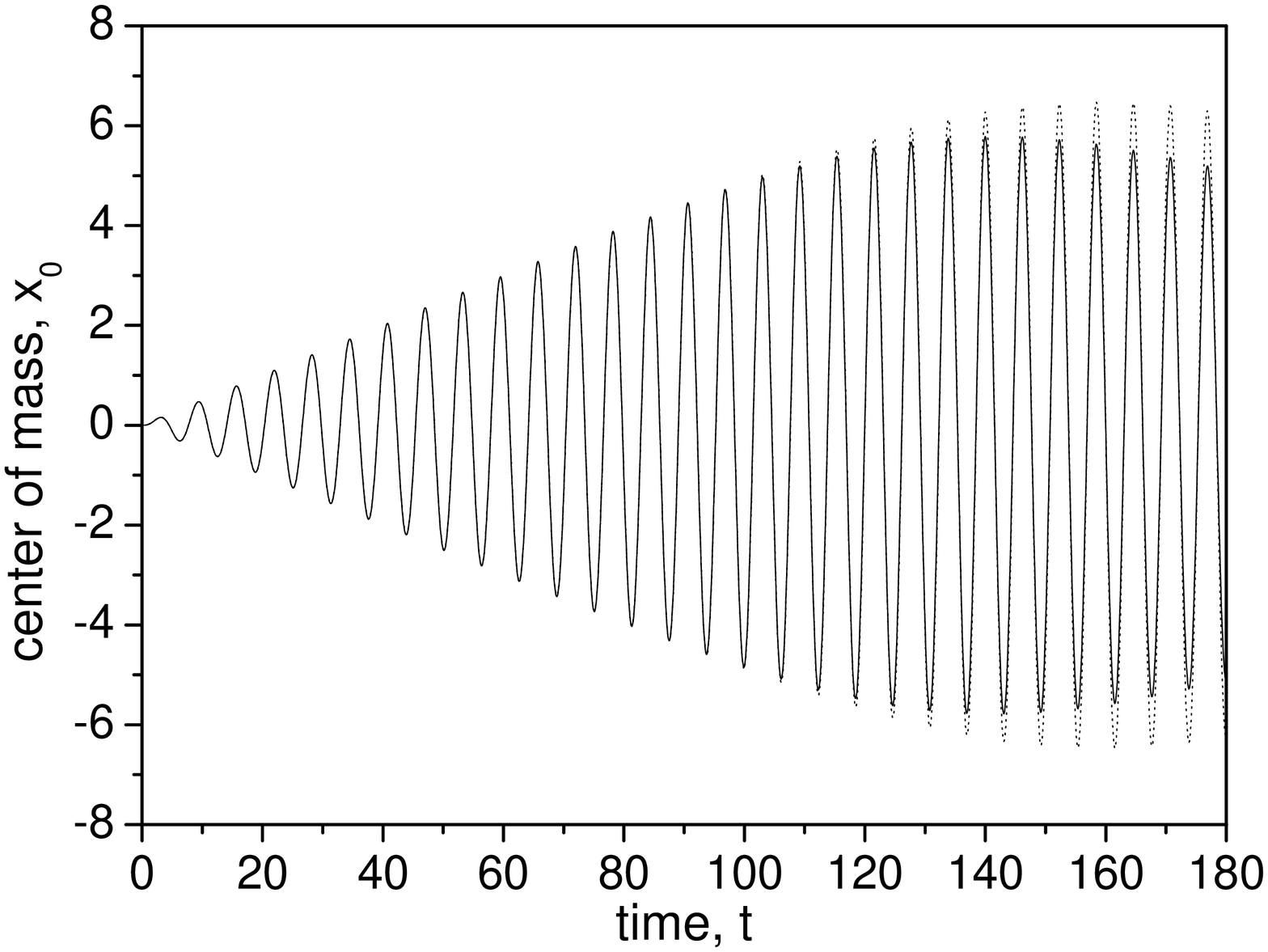}}
\caption{The width and center of mass oscillations of the repulsive BEC
in an anharmonic trap with the forced periodical oscillation of the
trap center with the frequency $w = w_x$ and $w_{\eta}/w_x = 1.966$.
Nonlinear coefficient, g equals to 0.4. The parameters are $V_2 =
0.5$,\ $V_4 = 0.0005,\ h = 0.1$. Solid and dotted lines stand for PDE
and ODE simulations respectively.} \label{nores}
\end{figure}

To check this assertion we carried out ODE and PDE simulations when
$w_{\eta}/w_x \neq 2$. In figure \ref{nores}, double resonance in
oscillations of the repulsive condensate width and center of mass is
presented when the nonlinearity coefficient $g = 0.4$. In this case the
ratio $w_{\eta}/w_x = 1.966$. Here the parameters are $V_2 = 0.5$,\
$V_4 = 0.0005,\ h = 0.1$. The value of the forced oscillation is taken
$w = w_x$. In spite of that the ratio of eigenfrequencies of the BEC
width and center of mass oscillation is not equal to 2, one can observe
double resonance in the oscillations.

Simulations of double resonance presented in figures \ref{figcenter}
and \ref{nores} relate to the case of the repulsive condensate. Let us
now consider an attractive condensate where the matter wave solitons
can exist. As seen from figure \ref{wawx} for the attractive BEC ($g <
0$) the ratio $w_{\eta}/w_x$ is not equal to 2 and exact resonance is
impossible in this case. Nevertheless the ratio $w_{\eta}/w_x$ remains
to be close to 2 (in considered range of the BEC parameters) and one
can expect resonant behavior of the width and the center of mass
oscillations of the attractive BEC under forced oscillations of the
trap potential position with the frequency $w = w_x$.

\begin{figure}[h]
\centerline{\includegraphics[width=8cm,angle=-90]{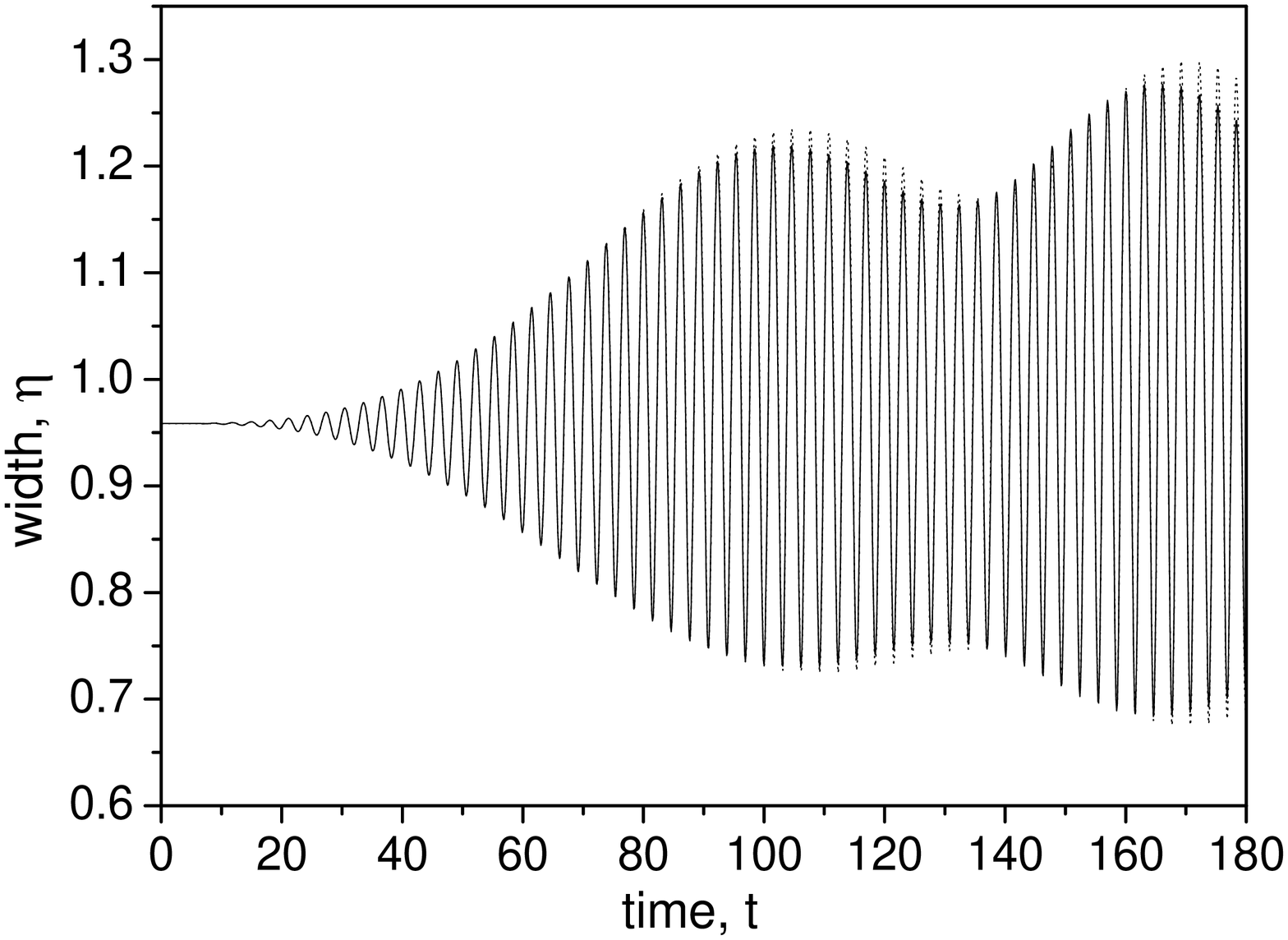}}
\centerline{\includegraphics[width=8cm,angle=-90]{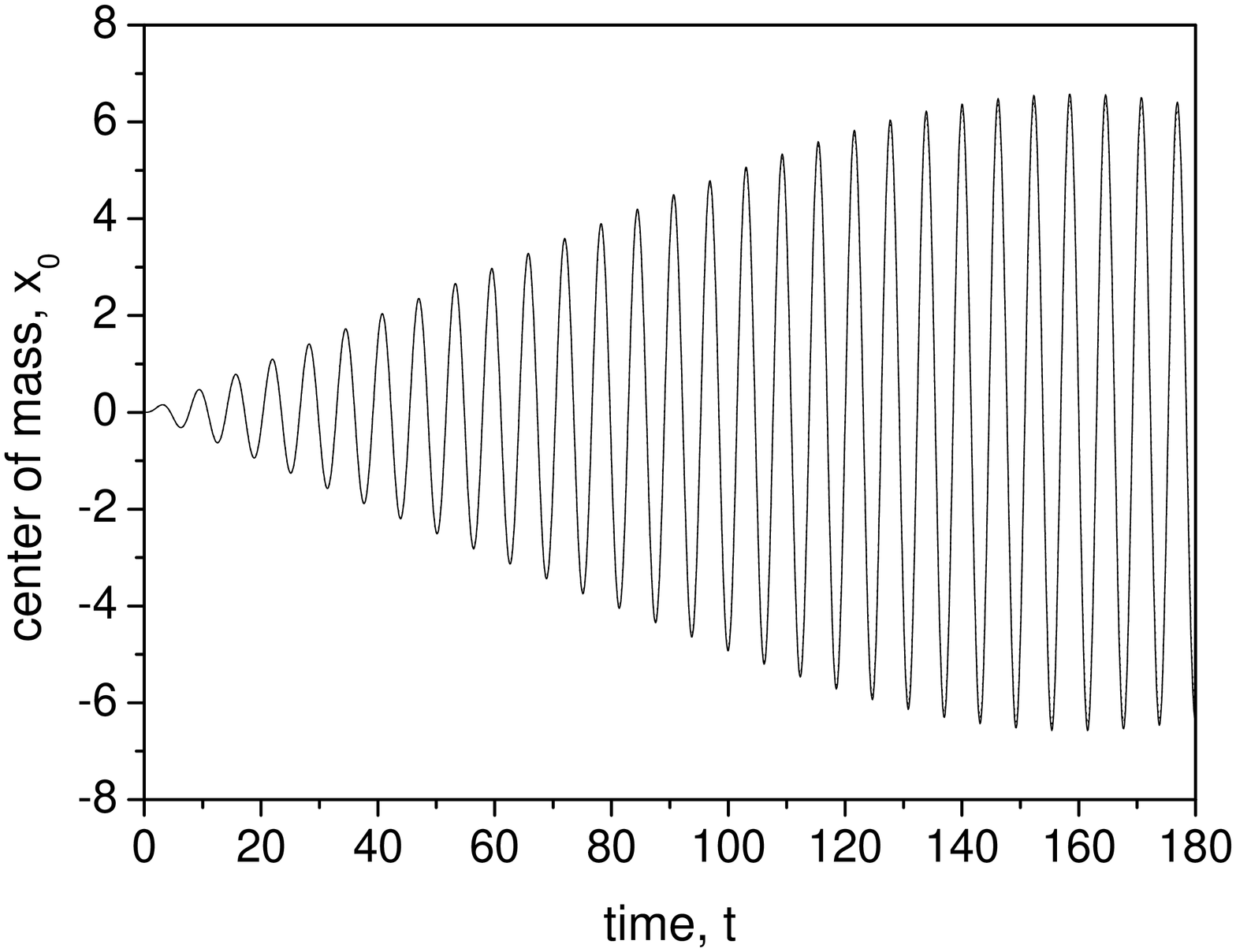}} \caption{
The width and center of mass oscillations of the attractive BEC in an
anharmonic trap with the forced periodical oscillation of the trap
center with the frequency $w = w_x$ and $w_{\eta}/w_x =2.046$.
Nonlinear coefficient, $g$ equals to -0.4. The parameters are $V_2 =
0.5$,\ $V_4 = 0.0005,\ h = 0.1$. Solid and dotted lines stand for ODE
and PDE simulations respectively.} \label{nores1}
\end{figure}

In figure \ref{nores1} resonant behavior of oscillations of the
attractive condensate width and center of mass is depicted when the
nonlinearity coefficient $g = -0.4$. In this case the ratio
$w_{\eta}/w_x = 2.046$. Here the parameters are $V_2 = 0.5$,\ $V_4 =
0.0005,\ h = 0.1$. The value of the forced oscillation is taken $w =
w_x$. As in the case of {\it repulsive} BEC one can observe that the
oscillations close to double resonance in the case of {\it attractive}
BEC.

As seen the results of ODE simulations of the double resonance are in a
good agreement with full PDE ones at times $t < 100$ and then begins to
differ at larger times. At these times the amplitude of oscillations of
the BEC center of mass $x_0$ is great and anharmonic part of the trap
potential becomes noticeable that leads to difference between ODE and
PDE simulations.

\section{Conclusion}
In this paper we have studied collective oscillations of a quasi- one-
dimensional Bose gas in an ahharmonic trap under periodic oscillations
of the trap position in time. To describe evolution of oscillations we
use variational approach with Gausiian ansatz. Double resonance in the
condensate oscillations has been studied. Analytical expressions have
been derived for the growth of oscillation amplitudes in the resonance.

Analysis of the variational equations has shown existence of a double
resonance in oscillations of the center of mass and the width under
forced oscillations of the trap center position, provided that the
ratio of the eigenfrequencies $w_\eta/w_x = 2$ and the forced trap
position oscillation frequency $w=w_x$. It is shown that for a wide
range of values of the BEC parameters the ratio $w_\eta/w_x$ is close
to 2 and the behavior of the oscillations is close to resonant both for
repulsive and attractive Bose gases.

Theoretical predictions are confirmed by full numerical simulations of
the 1D GP equation.

\ack
F. Kh. A. is gratiful to FAPESP for partial support of his work.

\end{document}